\pdfoutput=1
\documentclass[aps, preprint, superscriptaddress, showpacs]{revtex4}

\preprint{RIKEN-iTHEMS-Report-25,UT-HET-145}

\usepackage[T1]{fontenc}
\usepackage{lmodern}
\usepackage{microtype}
\usepackage{amsmath,amssymb}
\usepackage{graphicx}
\usepackage{tikz}
\usepackage[utf8]{inputenc}
\usepackage{xcolor}
\usepackage{siunitx}
\usepackage{booktabs}
\usepackage{comment}
\usepackage[caption=false]{subfig}
\usepackage[pdftex,unicode]{hyperref}

\DeclareSIUnit\barn{b}

\begin{document}
%%%%%%%%%%%%%%%%%%%%%%%%%%%%%%%%%%
%%%%%%%%%%% Title page %%%%%%%%%%%
%%%%%%%%%%%%%%%%%%%%%%%%%%%%%%%%%%

%%%%%%%%%%%%%%%%%%%%%%%%%%%%%%%%%%
\title{Classifying extended Higgs models through the trilinear Higgs boson coupling measurement at future colliders}
%%%%%%%%%%%%%%%%%%%%%%%%%%%%%%%%%%
\author{Nagisa Hiroshima}
\affiliation{Department of Physics, Faculty of Engineering Science, Yokohama National University, Yokohama 240–8501, Japan}
\affiliation{RIKEN Center for Interdisciplinary Theoretical and Mathematical Sciences(iTHEMS), RIKEN, Wako 351-0198, Japan}
\affiliation{Department of Physics, University of Toyama, 3190 Gofuku, Toyama 930-8555, Japan}
%%%%%%%%%%%%%%%%%%%%%%%%%%%%%%%%%%
\author{Mitsuru Kakizaki}
\affiliation{Department of Physics, University of Toyama, 3190 Gofuku, Toyama 930-8555, Japan}
%%%%%%%%%%%%%%%%%%%%%%%%%%%%%%%%%%
\author{Shuhei Ohzawa}
\affiliation{Department of Physics, University of Toyama, 3190 Gofuku, Toyama 930-8555, Japan}
%%%%%%%%%%%%%%%%%%%%%%%%%%%%%%%%%%
\date{October 24, 2025}
%%%%%%%%%%%%%%%%%%%%%%%%%%%%%%%%%%
\begin{abstract}

We investigate the trilinear Higgs boson coupling derived from the functional forms of various extended Higgs potentials.
In light of experimental constraints on Higgs boson couplings, we focus on extended Higgs models in which the trilinear Higgs boson coupling is predominantly determined by the Standard Model (SM) Higgs field.
Such models include the nearly aligned Higgs effective field theory, classically scale-invariant models, pseudo-Nambu-Goldstone boson scenarios, tadpole-induced models, and others.
We also consider higher-order corrections, including top quark and new particle contributions that are often neglected, and discuss their impact on the trilinear Higgs boson coupling.
Finally, we show to what extent the functional forms of the Higgs potentials can be probed at future colliders.

\end{abstract}
%%%%%%%%%%%%%%%%%%%%%%%%%%%%%%%%%%
\maketitle

% \tableofcontents

%%%%%%%%%%%%%%%%%%%%%%%%%%%%%%%%%%%%%%%%%%%%%%%%%%%%%%%%%%%%%%%%%%%%
\section{Introduction}
%%%%%%%%%%%%%%%%%%%%%%%%%%%%%%%%%%%%%%%%%%%%%%%%%%%%%%%%%%%%%%%%%%%%

The Standard Model (SM) of particle physics has been well established as a low-energy effective theory valid up to $\mathcal{O}(100) \, \unit{\giga \electronvolt}$, following the discovery of the Higgs boson $h$ at the CERN Large Hadron Collider (LHC) in 2012~\cite{ATLAS:2012yve, CMS:2012qbp}.
So far, no collider signature has been observed that clearly contradicts the SM with a minimal Higgs sector.
The predicted relationships between the masses of elementary particles and their coupling strengths to the Higgs boson have been consistent with the data obtained at the LHC~\cite{ATLAS:2022vkf}.
However, new physics beyond the SM (BSM) is still necessary.
One reason is that theoretical puzzles remain in the Higgs sector:
There is no guiding principle for formulating the Higgs sector Lagrangian.
The widely accepted SM Higgs sector is based on the assumption of minimality, but nature may allow for a more complex structure.
The Higgs boson itself may not be an elementary particle,  but rather a composite one.
The dynamics underlying electroweak symmetry breaking is yet to be understood.
In addition to these issues in the Higgs sector, there remain phenomena that require BSM physics.
These include the baryon asymmetry of the Universe, the existence of dark matter, the tiny but non-zero neutrino masses, cosmic inflation, and so on.

To address the above-mentioned issues, many BSM scenarios have been proposed and discussed in the literature.
One approach is to pursue the so-called hierarchy problem, which arises from quadratically divergent quantum corrections to the squared mass of the Higgs boson.
If the cutoff scale is set at a high energy scale, such as the Planck scale, one must fine-tune the parameters unnaturally to reproduce the observed Higgs boson mass.
To tackle the hierarchy problem, various extensions of the Higgs sector have been considered. Supersymmetric models~\cite{Nilles:1983ge, Haber:1984rc}, classically scale invariant (CSI) models~\cite{Bardeen:1995kv, Gildener:1976ih}, composite Higgs models~\cite{Kaplan:1983fs, Kaplan:1983sm} are part of examples. 
BSM models are now diverse, and efficient methods are needed to discriminate those models using experimental data.

It should be noted that the physical quantities related to the Higgs potential, which are currently available, are the vacuum expectation value (VEV), $v=246~\mathrm{GeV}$~\cite{MuLan:2010shf}, and the mass of the discovered Higgs boson $m_{h}^{}=125~\mathrm{GeV}$~\cite{ATLAS:2023oaq}.
These correspond to the position and the curvature of the local minimum of the Higgs potential.
This implies that only the shape near the local minimum is determined, while the global structure of the potential remains undetermined.
However, understanding the global shape is essential for unveiling the dynamics of the electroweak phase transition. 
If the electroweak phase transition is of the first order, the observed baryon asymmetry of the Universe may be explained via electroweak baryogenesis scenarios, and stochastic gravitational waves may be detectable in future GW observatories, such as LISA~\cite{Baker:2019nia} and DECIGO~\cite{Kawamura:2018esd}.
Therefore, studying the Higgs potential should provide important clues to physics beyond the Standard Model.

The first step toward determining the global structure of the Higgs potential is to measure the trilinear Higgs boson coupling, $\lambda_{hhh}^{}$.
In discussions of the trilinear Higgs boson coupling, it is conventional to introduce the ratio $\kappa_{\lambda}^{}$, defined as
\begin{align}
\label{eq:kappa_lambda}
    \kappa_{\lambda}^{}
    = \frac{\lambda_{hhh}^{}}{\lambda_{hhh}^{\mathrm{SM}}},
\end{align}
where $\lambda_{hhh}^{\mathrm{SM}}$ denotes the value predicted by the SM with the minimal Higgs sector.
The trilinear Higgs boson coupling has been probed via Higgs boson pair production at the LHC, but is not yet sufficiently constrained.
In Run 2 of the LHC, the ATLAS and CMS collaborations reported constraints of $- 0.4 < \kappa_{\lambda}^{} < 6.3$~\cite{ATLAS:2022jtk} and $- 1.24 < \kappa_{\lambda}^{} < 6.49$~\cite{CMS:2022dwd}, respectively, at \qty{95}{\percent} confidence level (C.L.).
Future collider experiments, such as the High-Luminosity LHC (HL-LHC) and the International Linear Collider (ILC), are expected to significantly tighten these constraints.
At the HL-LHC, it is projected that the ATLAS and CMS collaborations will achieve $0.5 < \kappa_{\lambda}^{} < 1.6$ and~\cite{ATLAS:2022faz} and $0.35 < \kappa_{\lambda}^{} < 1.9$, respectively, at \qty{68}{\percent}~C.L.~\cite{Cepeda:2019klc}.
In the upgraded ILC with a center-of-mass energy of $\sqrt{s}=1~\mathrm{TeV}$ and a luminosity of $L=3~\mathrm{ab}^{-1}$, the trilinear Higgs boson coupling is expected to be measured with approximately \qty{10}{\percent} accuracy around the SM value $\kappa_{\lambda}^{} = 1$~\cite{ILCInternationalDevelopmentTeam:2022izu}.
In addition to these experiments, measurements of trilinear Higgs couplings are also anticipated at other future accelerator experiments. At the 100 TeV proton–proton collider (FCC-hh/SppC), with $L=30~\mathrm{ab}^{-1}$, the coupling precision is estimated to be
$\delta \kappa_{\lambda}\simeq 3.4\text{–}7.8\%$, and at the muon collider with
$\sqrt{s}=3~\mathrm{TeV}$ and $\mathcal{L}=2~\mathrm{ab}^{-1}$,
$\kappa_{\lambda}\approx 1\pm 0.15$ ($68\%$ C.L.) is expected~\cite{DiMicco:2019ngk, Accettura:2023ked}.
Other colliders, such as the photon collider and the future circular $e^+e^-$ collider, have also been explored in the literature for probing the Higgs self-coupling. 
For recent analyses, see Refs.~\cite{Berger:2025ijd, Maura:2025rpe}.

Under these circumstances, this paper discusses the classification of the functional form of the Higgs potential in order to obtain insights into the underlying physics.
Related work can be found in, for example, Refs.~\cite{Agrawal:2019bpm, Bahl:2023eau, Bahl:2025wzj}.
Ref.~\cite{Agrawal:2019bpm} demonstrates that double-Higgs production at the 27 TeV LHC and at a \qty{100}{\tera \electronvolt} proton-proton collider is useful for discriminating shapes of the Higgs potentials.
Also, Refs.~\cite{Bahl:2023eau,Bahl:2025wzj} aim to provide precise theoretical predictions for the trilinear Higgs boson coupling in renormalizable models,
including higher-order loop corrections and the dependence on the external momenta.
Since radiative corrections to the trilinear Higgs boson coupling can exceed \qty{10}{\percent} compared to the tree-level prediction, such corrections cannot be neglected given the expected precision of future collider experiments.
In this paper, we extend the analysis of Ref.~\cite{Agrawal:2019bpm} to a broader class of models, including those with non-renormalizable Higgs potentials, and incorporate certain loop corrections that were not considered in the previous work.
We then investigate the feasibility of probing the trilinear Higgs boson coupling at future collider experiments in several extended Higgs models: the nearly aligned Higgs Effective Field Theory (naHEFT)~\cite{Kanemura:2021fvp, Kanemura:2022txx}, CSI models~\cite{Coleman:1973jx, Gildener:1976ih, Hashino:2015nxa}, pseudo-Nambu–Goldstone boson (pNGB) scenarios~\cite{Kaplan:1983fs, Kaplan:1983sm}, and tadpole-induced models~\cite{Galloway:2013dma, Chang:2014ida}.

This paper is organized as follows. In Sec.~\ref{sec:extented-models}, we derive the trilinear Higgs boson couplings in the extended Higgs models, including one-loop corrections arising from the top quark and new particles.
Next, in Sec.~\ref{sec:feasibility}, we discuss the feasibility of probing these couplings at future collider experiments. Sec.~\ref{sec:summary} is devoted to the summary.

%%%%%%%%%%%%%%%%%%%%%%%%%%%%%%%%%%%%%%%%%%%%%%%%%%%%%%%%%%%%%%%%%%%%
\section{Higgs Potentials in Extended Models}
\label{sec:extented-models}
%%%%%%%%%%%%%%%%%%%%%%%%%%%%%%%%%%%%%%%%%%%%%%%%%%%%%%%%%%%%%%%%%%%%

Here, we briefly review the shapes of the Higgs potential in various extended Higgs models.
It should be noted that certain classes of BSM models can be well approximated by the SM with an extended Higgs potential.
Among these, we focus on cases where the Higgs potential $V(\phi)$ can be effectively described by the single SM Higgs field $\phi$, {\it i.e.}, in the nearly aligned limit.
For concreteness, we consider Higgs potentials inspired by the following types of models: the naHEFT~\cite{Kanemura:2021fvp, Kanemura:2022txx}, CSI models~\cite{Coleman:1973jx, Gildener:1976ih, Hashino:2015nxa}, the pNGB models~\cite{Kaplan:1983fs, Kaplan:1983sm}, and the tadpole-induced models~\cite{Galloway:2013dma, Chang:2014ida}.
We include possible loop contributions to the Higgs potential from the top quark and hypothetical particles, and classify the resulting Higgs potentials according to their functional forms.
Contributions from weak gauge bosons are much smaller than those from the top quark, hence are safely neglected in our analysis.
Most of the following discussion is based on the Higgs potential in the broken phase.
From the vacuum stability condition, 
\begin{align}
    \left. \frac{\partial V(\phi)}{\partial \phi} \right|_{\phi=v} = 0,
\end{align}
and the mass of the discovered Higgs boson $m_h^{}$ is obtained as
\begin{align}
    \left. \frac{\partial^2 V(\phi)}{\partial \phi^2} \right|_{\phi=v} = m_h^2.
\end{align}
The trilinear Higgs boson coupling is determined by taking the third derivative, 
\begin{align}
    \left. \frac{\partial^3 V(\phi)}{\partial \phi^3} \right|_{\phi=v} = \lambda_{hhh}.
\end{align}
We use the effective potential method, and the external momentum dependence of the trilinear Higgs boson coupling is neglected in our analysis.

%%%%%%%%%%%%%%%%%%%%%%%%%%%%%%%%%%%%%%%%%%%%%%%%%%%%%%%%%%%%%%%%%%%%
\subsection{Standard Model with the minimal Higgs sector}
%%%%%%%%%%%%%%%%%%%%%%%%%%%%%%%%%%%%%%%%%%%%%%%%%%%%%%%%%%%%%%%%%%%%

In the SM with the minimal Higgs sector, the potential at the one-loop level is expressed in the following form,
\begin{align}
\label{eq:SM_potential}
    V(\phi)
    = - A \phi_{}^{2} + B \phi_{}^{4} + C \phi_{}^{4} \ln \frac{\phi_{}^{2}}{Q_{}^{2}},
\end{align}
where $A$, $B$ and $C$ are parameters, and $Q$ denotes the renormalization scale.
Notice that the shift of $Q$ and Yukawa couplings in the logarithm are absorbed by redefining the parameters $B$.
Using the vacuum stability condition and the definition of the Higgs boson mass, two of the parameters are replaced by $v$ and $m_h$. 
Then we obtain the Higgs boson coupling as
\begin{align}
    \lambda_{hhh}^{}
    = \frac{3 m_h^2}{v} \left( 1 + \frac{16 C}{3} \frac{v_{}^{2}}{m_h^2} \right).
\end{align}
The coefficient $C$ results from loop diagrams, and is expected to be smaller than $B$.
In the SM, the coefficient $C$ is dominated by top quark loop contributions: 
\begin{align}
    C_\mathrm{SM} = - \frac{3 m_t^4}{16 \pi^2 v^4},
\end{align}
and thus the trilinear Higgs boson coupling is reduced to
\begin{align}
    \lambda_{hhh}^{\mathrm{SM}}
    = \frac{3 m_h^2}{v} \left( 1 - \frac{1}{\pi^2} \frac{m_t^{4}}{v^2 m_h^2} \right).
\end{align}
This value is used as a reference and as the denominator of the $\kappa$ parameter for each model throughout this paper.

%%%%%%%%%%%%%%%%%%%%%%%%%%%%%%%%%%%%%%%%%%%%%%%%%%%%%%%%%%%%%%%%%%%%
\subsection{Nearly aligned Higgs Effective Field Theory}
%%%%%%%%%%%%%%%%%%%%%%%%%%%%%%%%%%%%%%%%%%%%%%%%%%%%%%%%%%%%%%%%%%%%

One of the simplest extensions of the SM is such that the Higgs potential receives quantum corrections from additional heavy particles keeping the particle content at low energies is the same as the SM.
This type of the effective theory is called the nearly aligned Higgs effective field theory (naHEFT)~\cite{Kanemura:2021fvp,Kanemura:2022txx}.
The naHEFT Higgs potential in the broken phase has the form of
\begin{align}
    V = - A \phi^2 + B \phi^4 + \sum_i \xi_i [M_i^2 (\phi)]^2 \ln \frac{M_i^2(\phi)}{Q^2},
\end{align}
where $M_i^2(\phi)$ denotes the field-dependent mass squared of particle $i$ and $\xi_i^{}$ are loop-suppressed coefficients.
One example leading to the above potential is an inert scalar model.
Consider, in particular, an inert scalar model equipped with $N$ real scalar singlets with a common mass and coupling parameters~\cite{Kakizaki:2015wua}.
Namely, the scalar sector is invariant under an O($N$) symmetry.
After integrating out the inert scalars, the Higgs potential becomes
\begin{align}
    V = - A \phi^2 + B \phi^4 
    - \frac{3}{16 \pi^2} [M_t^2 (\phi)]^2 \ln \frac{M_t^2(\phi)}{Q^2}
    + \frac{N}{64 \pi^2} [M^2 (\phi)]^2 \ln \frac{M^2(\phi)}{Q^2},
\end{align}
where the field-dependent masses of the top quark and inert scalars are
\begin{align}
    M_t^2(\phi) = m_t^2 \frac{\phi^2}{v^2}, \quad 
    M^2(\phi) = \mu^2 + c \phi^2, 
\end{align}
respectively.
Here, $\mu$ is the common invariant mass of the inert scalars, and the coefficient $c$ arises from the coupling between the SM Higgs doublets and the inert scalars.
The resulting trilinear Higgs boson coupling is 
\begin{align}
    \lambda_{hhh}^{\mathrm{O}(N)} = 
    \frac{3 m_h^2}{v} \left( 1 - \frac{1}{\pi^2} \frac{m_t^{4}}{v^2 m_h^2} + \frac{N}{12 \pi^2} \frac{m_S^{4}}{v^2 m_h^2}\left( 1 - \frac{\mu^2}{m_S^2}\right)^3 \right),
\end{align}
where the common mass squared of the singlets is
\begin{align}
    m_S^2 = \mu^2 + c v^2.
\end{align}

The case where the invariant mass is zero, $\mu=0$, reduces to the potential of Eq.~(\ref{eq:SM_potential}).
In this case, the possible maximum deviation of $C_{\mathrm{O}(N)}$ from the SM prediction becomes
\begin{align}
    C_{\mathrm{O}(N)} = - \frac{3 m_t^4}{16 \pi^2 v^4} 
    + \frac{N m_S^4}{64 \pi^2 v^2}.
\end{align}
then 
\begin{align}
    \lambda_{hhh}^{\mathrm{O}(N)}
    = \frac{3 m_h^2}{v} \left( 1 - \frac{1}{\pi^2} \frac{m_t^{4}}{v^2 m_h^2} + \frac{N}{12 \pi^2} \frac{m_S^{4}}{v^2 m_h^2}\right).
\end{align}
On the other hand, if the invariant mass $\mu$ is larger than the electroweak scale, the expansion of the Higgs potential by $\phi^2/\mu^2$ or $\phi^2/m_S^2$ is valid and results in the shape of the Coleman-Weinberg potential with higher-dimensional operators, as we show below.

In the effective field theory (EFT) approach, the general Higgs potential in the broken phase is described by the so-called Higgs EFT while that in the unbroken phase by the SMEFT~\cite{Buchmuller:1985jz, Grzadkowski:2010es, Giudice:2007fh}
In such an EFT framework, contributions from new physics can be treated without specifying each model realization. 
In the decoupling limit, the Higgs potentials lead to the Landau-Ginzburg potential with higher-dimensional operators.
These are suppressed by powers of the new physics scale $\Lambda$.
Assuming $Z_{2}^{}$ symmetry in the Higgs potential, the dimension-five operator is eliminated, and in general, the dimension-six operator contributes dominantly among higher-dimensional operators. 
Taking into account the loop contributions from non-decoupling particles, such as the top quark, the Higgs potential with the dimension-six operator at the one-loop level results in
\begin{align}
\label{eq:SMEFT_potential}
    V(\phi)
    = - A \phi_{}^{2} + B \phi_{}^{4} + C \phi_{}^{4} \ln \frac{\phi_{}^{2}}{Q_{}^{2}} + D \phi_{}^{6},
\end{align}
where $D$ is a parameter that is inversely proportional to $\Lambda^2$.
The corresponding trilinear Higgs coupling is given by
\begin{align}
\label{eq:SMEFT_trilinear}
    \lambda_{hhh}^{\mathrm{dim}6}
    = \frac{3}{v} \left[m_{h}^{2} + \frac{16}{3} \left(C + 3 D v_{}^{2}\right) v_{}^{2}\right].
\end{align}
If all the new particles exhibit decoupling behavior, $C=C_\mathrm{SM}$, then Eq.~(\ref{eq:SMEFT_trilinear}) reduces to
\begin{align}
    \lambda_{hhh}^{\mathrm{dim}6} = \lambda_{hhh}^{\mathrm{SM}} + 48 D v_{}^{3}.
\end{align}
In terms of the SMEFT, the parameter $D$ is matched as $D=c_6 \lambda/(8 \Lambda^2)$.
In the $O(N)$ model described above, the Higgs potential is expanded by the powers of $c \phi_{}^{2} / \mu_{}^{2}$, leading to $D = N c_{}^{3} / 64 \pi_{}^{2} \mu_{}^{2}$.
The ratio is $\kappa_{\lambda}^{\mathrm{dim6}} =\lambda_{hhh}^{\mathrm{dim6}}/\lambda_{hhh}^{\mathrm{SM}}\simeq 1.1$, taking $D = 0.1~\unit{\TeV}^{-2}$ as a benchmark case.

%%%%%%%%%%%%%%%%%%%%%%%%%%%%%%%%%%%%%%%%%%%%%%%%%%%%%%%%%%%%%%%%%%%%
\subsection{Classical Scale Invariance (CSI) Model}
\label{sebsec:CSI}
%%%%%%%%%%%%%%%%%%%%%%%%%%%%%%%%%%%%%%%%%%%%%%%%%%%%%%%%%%%%%%%%%%%%

The idea of classical scale invariance (CSI) was originally introduced by Bardeen as a paradigm to address the hierarchy problem~\cite{Bardeen:1995kv}. 
In CSI models, the Lagrangian contains no explicit mass term due to scale invariance, and the Higgs mass term is generated dynamically via the Coleman–Weinberg mechanism~\cite{Coleman:1973jx}, which triggers electroweak symmetry breaking. 
The minimal CSI model with a single Higgs doublet has already been experimentally excluded~\cite{Coleman:1973jx}, and phenomenologically viable CSI models necessarily involve extended scalar sectors~\cite{Gildener:1976ih}. 
Based on this framework, many models of electroweak symmetry breaking have been proposed and extensively studied. 
These models exhibit characteristic features a universal enhancement of the trilinear Higgs coupling as large as 70\% at the leading order~\cite{Hashino:2015nxa}. 
They also predict a strongly first-order electroweak phase transition~\cite{Hashino:2016rvx}. The associated stochastic gravitational wave spectrum depends sensitively on the number and masses of extra fields, and thus provides a unique probe of the Higgs potential and the underlying dynamics. 
In this context, the combined information from future measurements of the trilinear Higgs coupling and gravitational wave signals be important for discriminating among extended Higgs sectors. 
For concreteness, we focus on CSI models with $N$ additional isospin-singlet scalars which are subject to a global $O(N)$ symmetry and discuss how these can be distinguished from similar scenarios by achieving the synergy of collider and gravitational wave observations.

In the CSI models, there is a flat direction in the Higgs potential at tree level.
Although the tree-level Higgs potential vanishes at tree level, the Higgs potential is lifted due to quantum effects and takes the form of 
\begin{align}
\label{eq:CSI_potential}
    V(\phi)
    = A \phi_{}^{4} + B \phi_{}^{4} \ln \frac{\phi_{}^{2}}{Q_{}^{2}},
\end{align}
where \textit{A} and \textit{B} are parameters characterizing the models and $Q$ is a renormalization scale.
In the CSI model, the parameter $B$ is proportional to the fourth power of the mass of the particle in the loop.
The trilinear Higgs coupling at the one-loop level is obtained as
\begin{align}
\label{eq:CSI_trilinear}
    \lambda_{hhh}^{\mathrm{CSI}}
    = \frac{5 m_{h}^{2}}{v},
\end{align}
That is, the trilinear Higgs coupling $\lambda_{hhh}^{\mathrm{CSI}}$ is a constant and is independent of the parameters of the CSI model.
The predicted trilinear Higgs boson coupling is larger than that of the SM counterpart by around 80\%.

%%%%%%%%%%%%%%%%%%%%%%%%%%%%%%%%%%%%%%%%%%%%%%%%%%%%%%%%%%%%%%%%%%%%
\subsection{Pseudo-Nambu-Goldstone Boson  (pNGB) Model}
\label{sebsec:pNGB}
%%%%%%%%%%%%%%%%%%%%%%%%%%%%%%%%%%%%%%%%%%%%%%%%%%%%%%%%%%%%%%%%%%%%

One well-known example of composite Higgs models is the pseudo-Nambu--Goldstone boson (pNGB) model, originally proposed in Refs.~\cite{Kaplan:1983fs, Kaplan:1983sm}.  
In this framework, the electroweak sector of the SM is embedded in a subgroup $H$, which emerges at a scale $f \sim \mathcal{O}(1)\,\unit{\tera\electronvolt}$ through the dynamical breaking of a global symmetry group $G$.  
This construction is analogous to the origin of the pion mass in QCD, which is protected against quantum corrections.  
The physical observables in such models depend on the ratio of the VEV $v$ to the symmetry-breaking scale $f$, conventionally parameterized as $\xi = v^2/f^2$.  
Quantum corrections are meaningful only after symmetry breaking, and the mechanism thereby provides a potential solution to the hierarchy problem.  
If a large hierarchy exists ($v \ll f$), the pNGB sector is effectively decoupled from the SM sector and its contributions become negligible.  
Conversely, if no large hierarchy is present ($v \sim f$), the pNGB model reduces to the minimal technicolor scenario~\cite{Hill:2002ap, Lane:2002wv, Chivukula:2000mb}.  
In general, consistency with measurements at the Large Electron-Positron Collider (LEP) requires $\xi < 0.1$~\cite{Bellazzini:2014yua}.

The Higgs potential in pNGB models vanishes at tree level and is generated through the Coleman--Weinberg mechanism at one-loop order~\cite{Coleman:1973jx}.  
It can be expressed as a power series in trigonometric functions, whose functional form depends on the specific operators included in the model.  
Consequently, models containing higher-order terms lead to different expressions from those restricted to the lowest-order contributions.  
A characteristic feature of these models is that the gauge couplings between the Higgs boson and the weak gauge bosons are identical to those of the Minimal Composite Higgs Model (MCHM):  
\begin{align}
\label{eq:MCHM_hVV_coupling}
    g_{hVV}^{\mathrm{MCHM}}
    = g_{hVV}^{\mathrm{SM}} \sqrt{1 - \xi}.
\end{align}
The shape of the Higgs potential also depends on how the SM fermions are embedded into representations of $\mathrm{SO}(5)$.
For deviation patterns in Higgs coupling constants in various kinds of MCHMs, see Ref.~\cite{Carena:2014ria, Kanemura:2014kga}. 
In the following, we analyze two benchmark scenarios: the so-called MCHM4~\cite{Agashe:2004rs} and MCHM5~\cite{Contino:2006qr}.  
In the former, SM fermions transform as spinor representations of $\mathrm{SO}(5)$, whereas in the latter they are embedded in the fundamental representation of $\mathrm{SO}(5)$. These models also emerge from warped extra dimension scenarios.

%%%%%%%%%%%%%%%%%%%%%%%%%%%%%%%%%%%%%%%%%%%%%%%%%%%%%%%%%%%%%%%%%%%%
\subsubsection{MCHM4 at the lowest order}
%%%%%%%%%%%%%%%%%%%%%%%%%%%%%%%%%%%%%%%%%%%%%%%%%%%%%%%%%%%%%%%%%%%%

In MCHM4, the minimally truncated Higgs potential that can lead to successful EWSB takes the form
\begin{align}
\label{eq:MCHM4_potential}
    V(\phi)
    = A \cos \frac{\phi}{f} - B \sin^{2} \frac{\phi}{f},
\end{align}
where $A$ and $B$ are positive coefficients determined by physics at the energy scale at which the perturbative description breaks down.
We hereafter refer to this potential as MCHM4$_{(0)}$.
The parameters in the potential is related to the parameter $\xi$ as 
\begin{align}
\label{eq:MHCM4_VEV}
    \xi
    = \frac{v^{2}}{f^{2}}
    = \sin^{2} \frac{\langle \phi \rangle}{f}
    = 1 - \left(\frac{A}{2B}\right)^{2}.
\end{align}
The trilinear Higgs coupling at one-loop order is then given by
\begin{align}
\label{eq:MCHM4_trilinear}
    \lambda_{hhh}^{\mathrm{MCHM4}_{(0)}}
    = \lambda_{hhh}^{\mathrm{SM, tree}} \sqrt{1 - \xi}.
\end{align}
Note that the dependence on $\xi$ is the same as that shown in Eq.~(\ref{eq:MCHM_hVV_coupling}).
A similar dependence appears in the Yukawa couplings to fermions:
\begin{align}
\label{eq:MCHM4_yukawa}
    g_{hff}^{\mathrm{MCHM4}}
    = g_{hff}^{\mathrm{SM}} \sqrt{1 - \xi}.
\end{align}
Exploiting this property, the trilinear Higgs coupling at lowest order can be indirectly probed through precision measurements of the bottom-quark Yukawa coupling, which are planned at an early stage of the ILC program.  

%%%%%%%%%%%%%%%%%%%%%%%%%%%%%%%%%%%%%%%%%%%%%%%%%%%%%%%%%%%%%%%%%%%%
\subsubsection{MCHM4 with higher-order corrections}
%%%%%%%%%%%%%%%%%%%%%%%%%%%%%%%%%%%%%%%%%%%%%%%%%%%%%%%%%%%%%%%%%%%%

When higher-order contributions are taken into account, additional terms appear in addition to Eq.~(\ref{eq:MCHM4_potential}).
In particular, we consider the case where the Higgs potential is expressed as~\cite{Agashe:2004rs}
\begin{align}
\label{eq:MCHM4higher_potential}
    V(\phi)
    = A \cos \frac{\phi}{f}
    - B \sin^{2} \frac{\phi}{f}
    + C \sin^{4} \frac{\phi}{f},
\end{align}
where $A$, $B$, and $C$ are dimensionless coefficients.
We hereafter refer to this potential as MCHM4$_{(1)}$.
Eq.~(\ref{eq:MCHM4higher_potential}) can be regarded as an approximate extension of Eq.~(\ref{eq:MCHM4_potential}) in that the second term originates from fermion-loop contributions, which appear as powers of $\cos(\phi / f)$, while the third term arises from gauge boson loops, as in the lowest-order analysis.  
Since the Higgs potential in MCHM4 is generated radiatively, we expect that the coefficients satisfy the approximate hierarchy $A \gtrsim B \gtrsim C$.  

The vacuum misalignment $\xi$ and the mass squared of the Higgs boson are given by
\begin{align}
\label{eq:MCHM4higher_VEV&Mass}
    \xi
    \simeq \frac{A}{4C} \frac{1 + 2B/A}{1 + B/4C} + \mathcal{O}(\xi^{2}),
    \quad
    m_{h}^{2}
    = 2 v^2 [B/f^4 + 2C/f^4 (2 - 3 \xi)].
\end{align}
Here we assume that $\xi$ is sufficiently small.  
The numerator of $\xi$ may take either $(1 + 2B/A)$ or $(-1 + 2B/A)$; however, the latter is unphysical. It requires either a negative VEV $\langle \phi \rangle$ or a VEV larger than the scale $f$.  
Therefore, we restrict our discussion to the physical branch.  

Using the lowest-order expression of Eq.~(\ref{eq:MCHM4_trilinear}), the trilinear Higgs boson coupling at higher order can be written as
\begin{align}
\label{eq:MCHM4hihger_trilinear&ratio}
    \lambda_{hhh}^{\mathrm{MCHM4} (1)}
    = \lambda_{hhh}^{\mathrm{MCHM4} (0)}
       \left(1 - \frac{8 C \xi}{f_{}^{4}}\frac{v^{2}}{m_{h}^{2}} \right).
\end{align}
In the limit $C \to 0$, this expression reduces to Eq.~(\ref{eq:MCHM4_trilinear}).
Depending on the value of the parameter $C$, the trilinear Higgs coupling can deviate from that inferred from the SM fermion coupling, which necessitates direct measurements of this quantity.

\begin{figure}[t]
    \centering
    \includegraphics[width=0.5\linewidth]{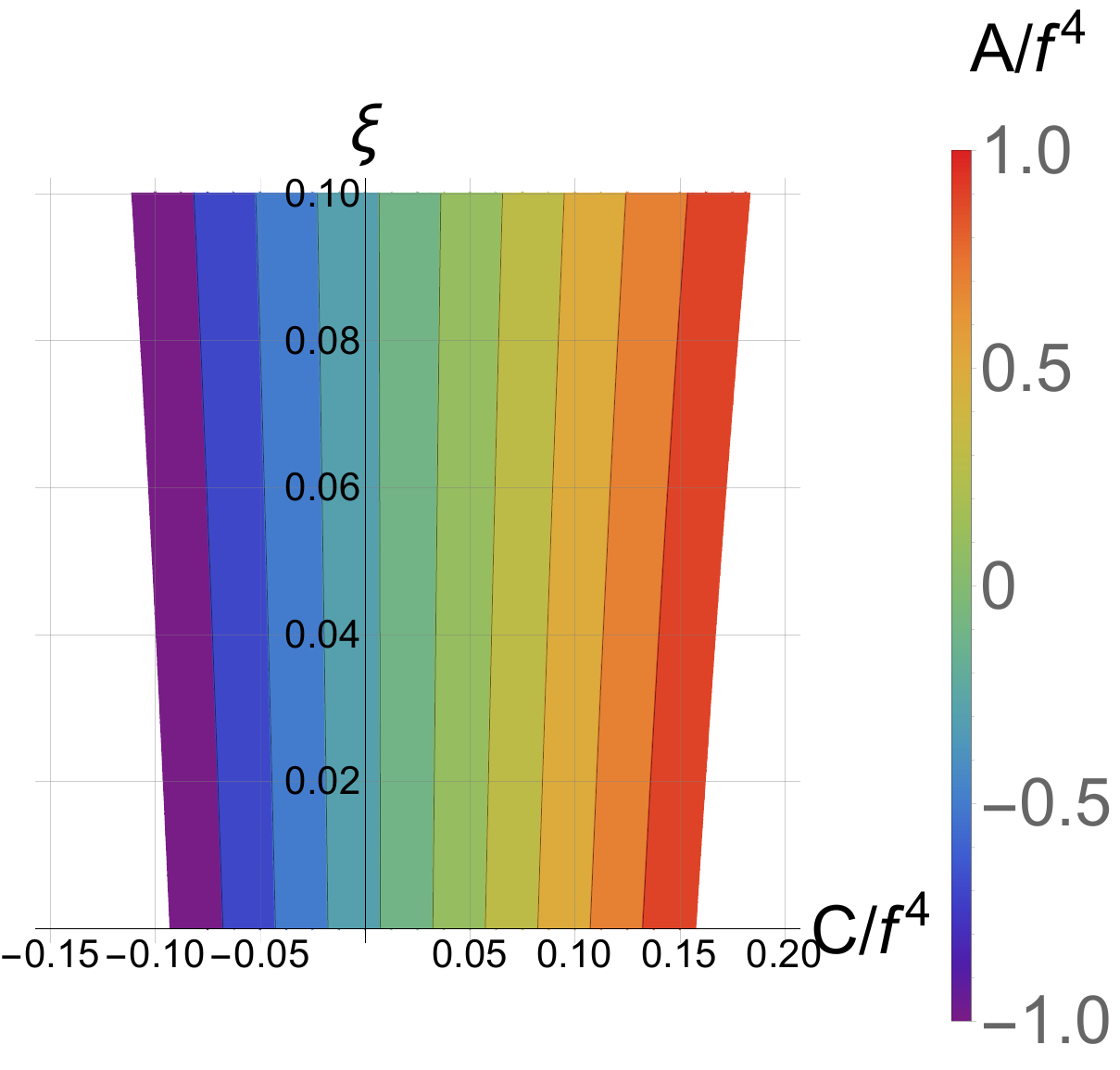}
    \caption{\footnotesize
    Contours of $A / f_{}^{4}$ obtained from the vacuum and mass conditions
    in Eq.~\ref{eq:MCHM4higher_VEV&Mass}.  
    In the perturbative regime for $A$, the range $-0.12 < C / f_{}^{4} < 0.19$ is allowed.
    }
    \label{fig:contour_alpha}
\end{figure}

Figure~\ref{fig:contour_alpha} shows the contours of $A/f^4$ in the $\xi$--$C / f_{}^{4}$ plane.  
We find that the allowed region $-0.12 < C / f_{}^{4} < 0.19$ narrows as $\xi$ decreases.  
To satisfy $\xi \ll 1$, approximately $A / 4C < 1$ is required.  
Since the signs of $A$ and $C$ are not constrained by the vacuum condition, both coefficients can in principle be negative.
Unless the ultraviolet completion is specified, the signs of these parameters are not fixed.

\begin{figure}[t]
    \centering
    \includegraphics[width=0.5\linewidth]{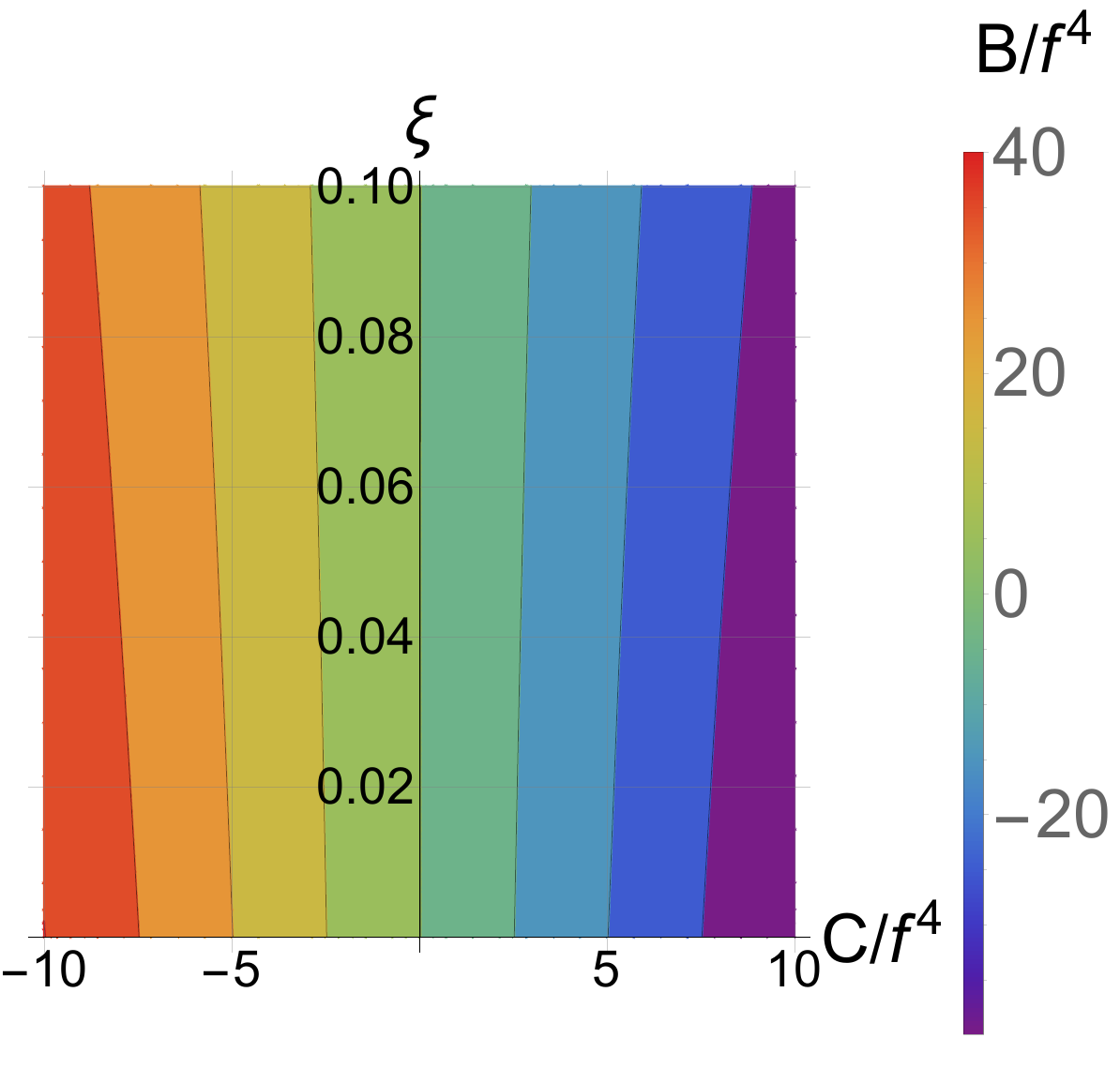}
    \caption{\footnotesize
    Contours of $B$ from the mass condition in Eq.~(\ref{eq:MCHM4higher_VEV&Mass}).  
    We consider the region where $\xi < 0.1$.  
    At $\xi = 0.1$, the magnitude of $B / f_{}^{4}$ must be roughly three times larger than that of $C / f_{}^{4}$.  
    For positive $B / f_{}^{4}$, one finds $C / f_{}^{4} \lesssim 0.04$.
    }
    \label{fig:contour_beta}
\end{figure}

The relation between $B$ and $C$ implied by the mass condition makes $\xi$ effectively dependent on the combination of $A$ and $C$.  
Since the Higgs boson mass squared depends on $B$, $C$, and $\xi$, some degrees of tuning are required, with the viable parameter space illustrated in Fig.~\ref{fig:contour_beta}.  
As seen in the figure, either $B$ or $C$ (or both) must be negative.  

\begin{figure}[t]
    \centering
    \includegraphics[width=0.8\linewidth]{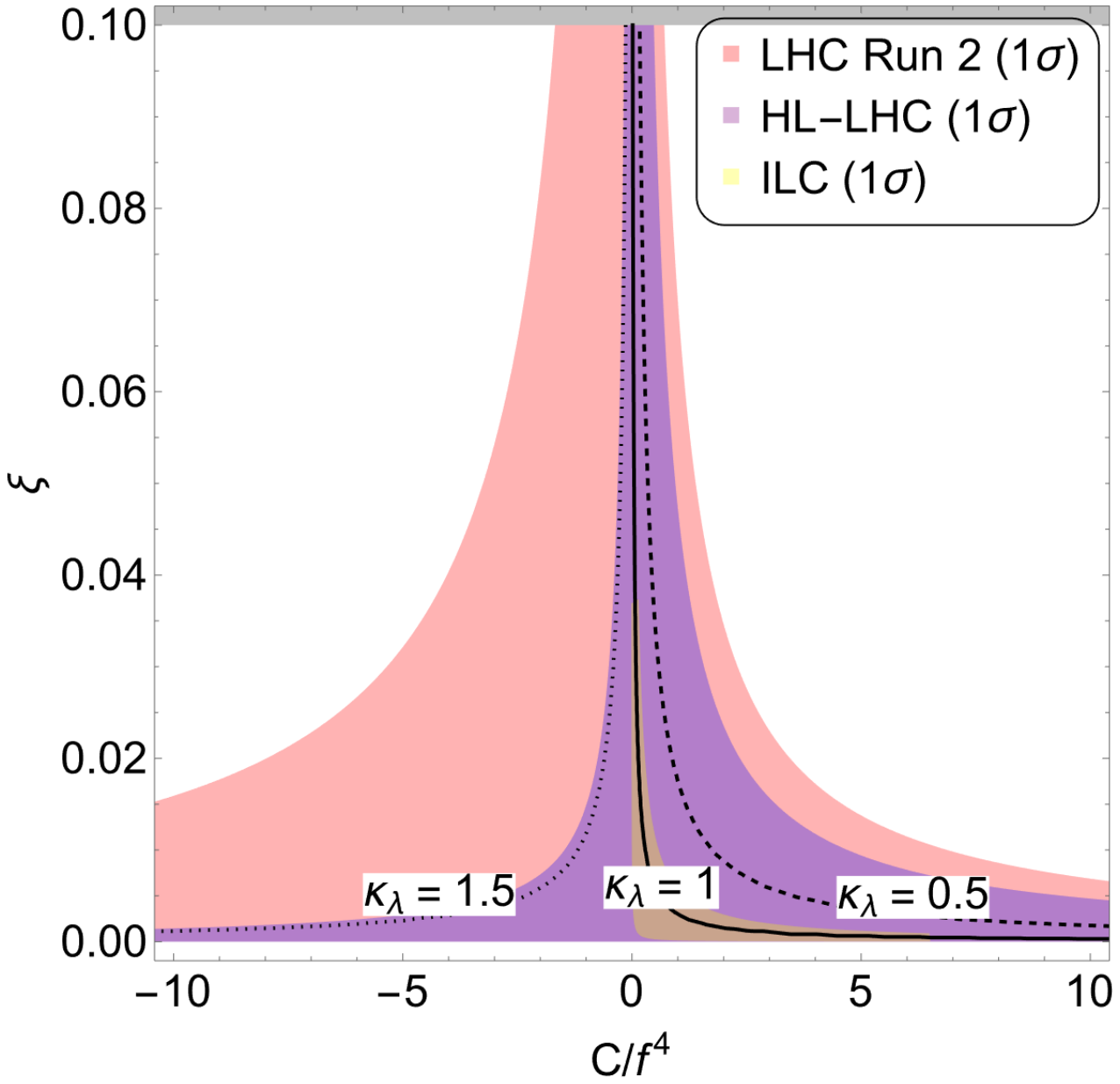}
    \caption{\footnotesize
        Allowed parameter region of MCHM4$_{(1)}$ including higher-order corrections, under the (expected) constraint on $\kappa_{\lambda}$ from the LHC and future collider experiments.
        The solid, dashed, and dotted contours represent the allowed regions in light of the results from the LHC Run 2, the HL-LHC, and the ILC with $\sqrt{s}=\qty{1}{\TeV}$.
        The gray-shaded area corresponds to the region where $\xi > 0.1$.
    }
    \label{fig:MCHM4ResultPlot}
\end{figure}

Figure~\ref{fig:MCHM4ResultPlot} shows the allowed parameter space of the MCHM4$_{(1)}$, where the trilinear Higgs boson coupling is evaluated including higher-order corrections. The vertical axis corresponds to $\xi$.
The horizontal axis represents the coefficient $C$ of the $\sin^4(h/f)$ term in the Higgs potential.
In MCHM4$_{(1)}$, $C/f^4$ is in generic an $\mathcal{O}(1)$ parameter so that a perturbative evaluation of the effective potential remains valid. 
Under this assumption, the magnitude of $C$ is moderately constrained. For example, at $\xi = 0.1$, we obtain $-0.15 \lesssim C/f^4 \lesssim 0.2$.
When $\xi$ becomes smaller, a wider range of $C$ values is allowed. However, excessively large values of $C$ would invalidate the perturbative treatment. 
Therefore, regions with $C/f^4 \gtrsim \mathcal{O}(10)$ are theoretically disfavored.

%%%%%%%%%%%%%%%%%%%%%%%%%%%%%%%%%%%%%%%%%%%%%%%%%%%%%%%%%%%%%%%%%%%%
\subsubsection{MCHM5 at the lowest order}
%%%%%%%%%%%%%%%%%%%%%%%%%%%%%%%%%%%%%%%%%%%%%%%%%%%%%%%%%%%%%%%%%%%%

In MCHM5, where the SM fermions are embedded in the fundamental representation of $\mathrm{SO}(5)$~\cite{Contino:2006qr}, the Higgs potential at lowest order takes the form
\begin{align}
\label{eq:MCHM5_potential}
    V(\phi) = - A \sin^{2} \frac{\phi}{f} + B \sin^{4} \frac{\phi}{f},
\end{align}
where parameters $A$ and $B$ take real values.  
In this case, radiative corrections from gauge bosons generate the potential, yielding a functional form proportional to $\sin^{2} (\phi / f)$. This is because 
interactions between fermions and Higgs fields require two particles of each type, meaning that the interaction terms must be proportional to $\sin_{}^{2}(\phi / f)$ or $\cos_{}^{2} (\phi / f)$.
Contributions from fermions to the Higgs potential takes the same functional form. 
The trilinear Higgs coupling is given by
\begin{align}
\label{eq:MCHM5_trilinear}
    \lambda_{hhh}^{\mathrm{MCHM5}}
    = \lambda_{hhh}^{\mathrm{SM, tree}} \frac{1 - 2 \xi}{\sqrt{1 - \xi}},
\end{align}
which exhibits the same dependence on $\xi$ as the Yukawa couplings:
\begin{align}
\label{eq:MCHM5_yukawa}
    g_{hff}^{\mathrm{MCHM5}}
    = g_{hff}^{\mathrm{SM}} \frac{1 - 2 \xi}{\sqrt{1 - \xi}}.
\end{align}

%%%%%%%%%%%%%%%%%%%%%%%%%%%%%%%%%%%%%%%%%%%%%%%%%%%%%%%%%%%%%%%%%%%%
\subsection{Tadpole-induced Model}
\label{subsec:tadpole}
%%%%%%%%%%%%%%%%%%%%%%%%%%%%%%%%%%%%%%%%%%%%%%%%%%%%%%%%%%%%%%%%%%%%

The composite Higgs sector can also be realized in so-called tadpole-induced models. This class of models was originally proposed in the context of superconformal technicolor~\cite{Azatov:2011ht, Azatov:2011ps}.
They contain an additional sector with auxiliary Higgs doublet fields that possess large quartic couplings but do not couple directly to SM quarks and leptons. 
In this framework, the mass parameter of the SM Higgs sector is positive, and thus spontaneous symmetry breaking does not occur on its own.  
Instead, electroweak symmetry breaking is triggered by a linear coupling between the SM Higgs and the auxiliary Higgs fields.  
This interaction induces a tadpole term in the Higgs potential, which effectively drives the symmetry breaking, resulting in distinctive signatures in the trilinear Higgs boson coupling~\cite{Galloway:2013dma, Chang:2014ida}.

Tadpole-induced models are characterized by the presence of an auxiliary Higgs field $\Sigma$, added to the SM Higgs doublet $\Phi$.  
Electroweak symmetry breaking is realized through a tadpole term that couples $\Phi$ and $\Sigma$.
Such a feature can be realized by the following Higgs potential:
\begin{align}
\label{eq:tadpole_potential}
    V(\Phi, \Sigma)
    = m_{H}^{2} \Phi_{}^{\dagger} \Phi + m_{\Sigma}^{2} \Sigma_{}^{\dagger} \Sigma - \kappa_{}^{2} \left(\Sigma_{}^{\dagger} \Phi + \mathrm{h.c.} \right) + \lambda_{\Sigma}^{} \left(\Sigma_{}^{\dagger} \Sigma \right)_{}^{2},
\end{align}
where the parameters $m_{H}^{2}$, $m_{\Sigma}^{2}$, $\kappa^2$ and $\lambda_\Sigma^{}$ are taken to be positive.  
The quartic term of the SM Higgs field, $\left(\Phi^{\dagger}\Phi\right)^{2}$, is assumed to be negligibly small so that electroweak symmetry breaking is induced by the tadpole term.  
We choose the gauge $\Phi = 1/\sqrt{2} (0,\, \phi)^{T}$, $ \quad \Sigma = 1/\sqrt{2} (0,\, \sigma)^{T}$ for the SM Higgs doublet and the auxiliary Higgs field, respectively. 
In the limit where the auxiliary Higgs field are decoupled, the mass of the auxiliary Higgs field reduces to $m_{\sigma}^{} = 2 \lambda_{\Sigma}^{} f_{}^{}$, where we denote the VEV of the auxiliary Higgs field as $\langle \sigma \rangle = f$.  
The effective potential for $\phi$ in this limit is obtained by integrating out $\sigma$.  

In contrast to previous studies, we consider radiative corrections to the Higgs potential.
Then, at the one-loop level, the Higgs potential is obtained as
\begin{align}
\label{eq:tadpole_one-loop_potential}
    V(\phi)
    = A \phi_{}^{2} - B \phi_{}^{} + C \phi_{}^{4} \ln \frac{\phi_{}^{2}}{Q_{}^{2}} + \lambda \phi_{}^{4},
\end{align}
where $A>0$, $B>0$, and $C$ are model dependent parameters, and $Q$ denotes the renormalization scale.
For simplicity, we assume that the mass scale of additional hypothetical particles is sufficiently higher than the electroweak scale, so that their contributions to the Higgs potential can be neglected, {\it i.e.}, $C=C_\mathrm{SM}^{}<0$.
This implies that, in order to lift the Higgs potential, we need to include a positive quartic coupling $\lambda$, which was not considered in previous studies.
Meanwhile, to preserve the idea of the tadpole potential, the magnitude of $\lambda$ should be much smaller than unity.
Therefore, fine-tuning of model parameters is necessary to some extent in this framework.

\begin{figure}
    \centering
    \includegraphics[scale=0.6]{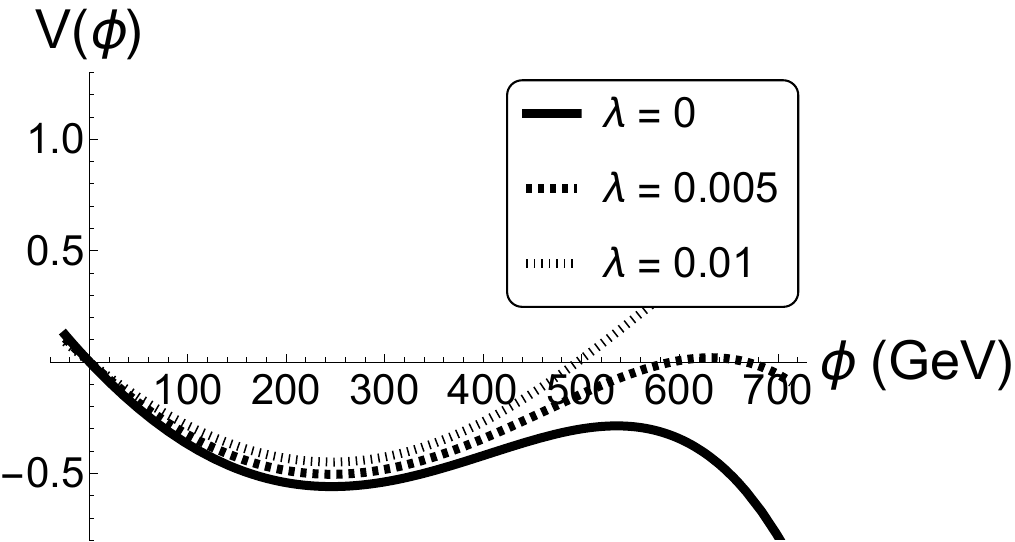}
    \caption{\footnotesize
    The potential shapes of the tadpole-induced model at the one-loop level, assuming $\lambda=0$ (solid), $0.005$ (dashed), and $0.1$ (dotted). 
    }
    \label{fig:tadpole}
\end{figure}

The trilinear Higgs boson coupling $\lambda_{hhh}^{\mathrm{tadpole}}$ is given by  
\begin{align}
\label{eq:tadpole_trilinear}
    \lambda_{hhh}^{\mathrm{tadpole}}
    = 4 v (13 C_\mathrm{SM} + 6 \lambda).
\end{align}
Suppose $\lambda \ll 1$, the trilinear Higgs boson coupling is predicted to be negative at the one-loop level.  
The potential of this model is illustrated in Fig.~\ref{fig:tadpole}.  
Since $\lambda$ should be kept small in the Higgs potential, the inclusion of radiative corrections can make the electroweak vacuum metastable.
Although higher-order corrections could potentially modify this result, a detailed analysis of vacuum stability is beyond the scope of the present study and will be discussed elsewhere.

%%%%%%%%%%%%%%%%%%%%%%%%%%%%%%%%%%%%%%%%%%%%%%%%%%%%%%%%%%%%%%%%%%%%
\section{Feasibility with future colliders}
\label{sec:feasibility}
%%%%%%%%%%%%%%%%%%%%%%%%%%%%%%%%%%%%%%%%%%%%%%%%%%%%%%%%%%%%%%%%%%%%

\begin{figure}[t]
    \begin{center}
    \includegraphics[scale=0.4]{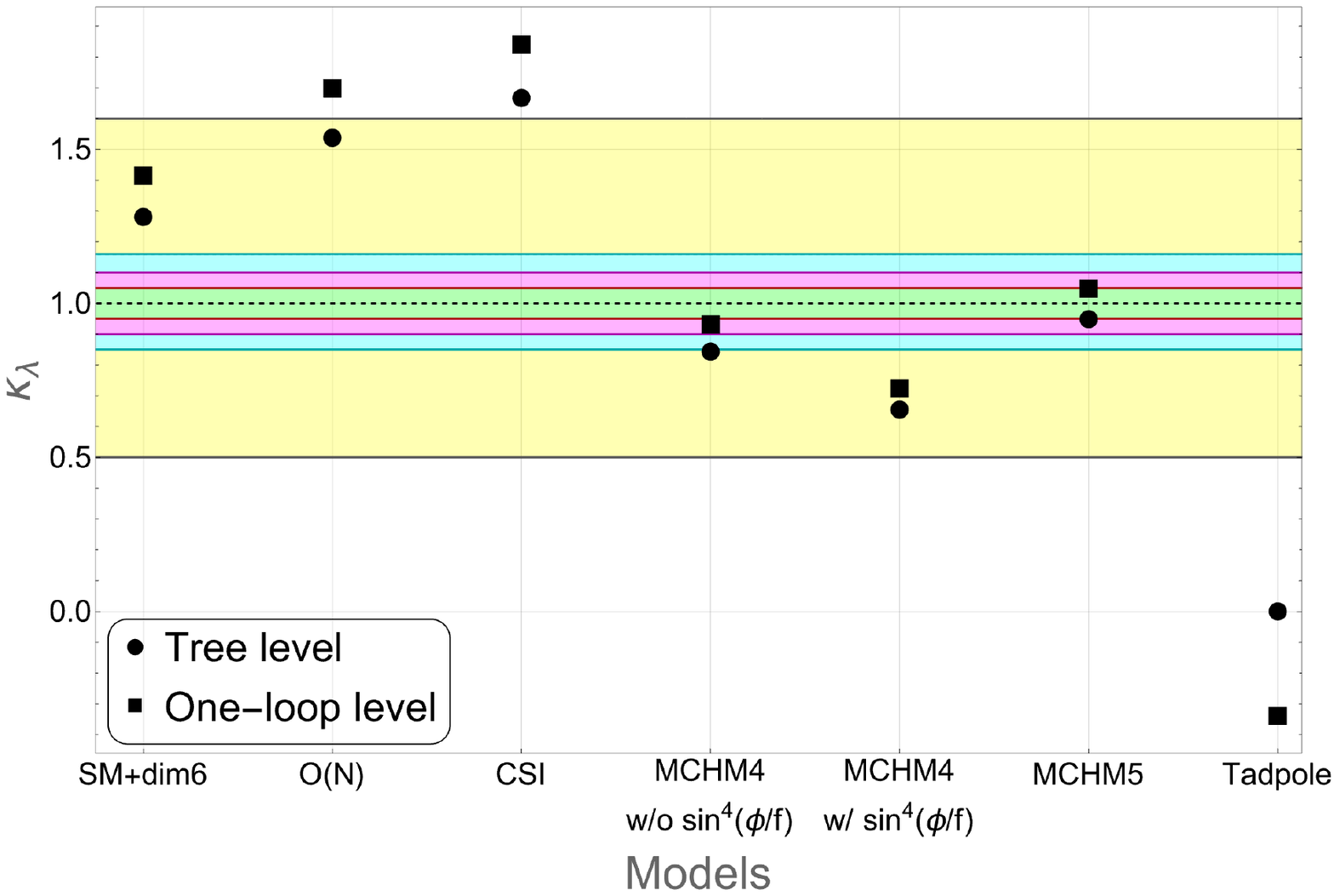}
    \end{center}
    \caption{\footnotesize
        Benchmark study on the one-loop predictions for the trilinear Higgs boson coupling $\kappa_\lambda$ in various extended Higgs models.
        Each point corresponds to a different model: the SM, extended models with dimension-six operators, the O($N$) scalar singlet model, the CSI model, MCHM4 without $\sin_{}^{4} (\phi / f)$ term, MCHM4 with $\sin_{}^{4} (\phi / f)$ term, MCHM5, and the tadpole-induced model (Tadpole), from left to right. 
        We show benchmark cases of $D = 0.1~\mathrm{TeV}_{}^{-2}$ for the dimension-six model, $N = 1$ and $m_{S}^{} = \qty{500}{\GeV}$ for the O($N$), $\xi = 0.1$ for MCHM4$_{(0)}$ and MCHM5, $(\xi, C / f_{}^{4}) = (0.1, \, 0.1)$ for MCHM4$_{(1)}$, $\lambda = 0$ for the tadpole-induced model.
        The shaded regions in yellow, cyan, magenta, and green indicate the $1 \sigma$ regions for $\kappa_{\lambda}^{\mathrm{exp}} = 1$ at HL-LHC (ATLAS), ILC \qty{1}{\TeV}, muon Collider, and \qty{100}{\TeV} $pp$ Collider, respectively.
    }
    \label{fig:kappa_lambda}
\end{figure}

Figure~\ref{fig:kappa_lambda} summarizes the results of our benchmark study on the one-loop predictions for the normalized trilinear Higgs boson coupling $\kappa_\lambda$ in various extended Higgs models.
Circles (squares) represent one-loop (tree level) values.  
The shaded regions indicate the projected $1\sigma$ sensitivities to $\kappa_\lambda = 1$ at future colliders: HL-LHC (ATLAS), ILC at $\sqrt{s} = 1~\mathrm{TeV}$, a muon collider, and a $100~\mathrm{TeV}$ proton--proton collider, shown in yellow, cyan, magenta, and purple, respectively. 

In the Higgs EFT framework, $\kappa_\lambda^{\mathrm{dim6}}$ depends on the parameter $D$. 
Since we truncate the potential at the dimension-six operator, the vacuum stability condition implies $D > 0$ and thus $\kappa_\lambda > 1$. 
We adopt $D = 0.1~\mathrm{TeV}_{}^{-2}$ as a benchmark.  
As shown in Fig.~\ref{fig:kappa_lambda}, this benchmark lies outside the HL-LHC sensitivity but within the $1\sigma$ reach of the ILC at $\sqrt{s} = 1~\mathrm{TeV}$, and can be detected at the $2\sigma$ level by both the muon and $100~\mathrm{TeV}$ proton--proton colliders.
A detailed study on how to constrain the trilinear Higgs boson coupling at the HL-LHC within the EFT framework is presented in Ref.~\cite{DiVita:2017eyz}.

In the $O(N)$ model, the trilinear Higgs boson coupling ratio $\kappa_\lambda^{\mathrm{O(N)}}$ depends on the parameters $N$ and $m_S$.  
We take $N = 1$ and $m_S^{} = 500~\mathrm{GeV}$ to illustrate our result.   
Future collider experiments at $\sqrt{s} = 1~\mathrm{TeV}$ and beyond can probe such benchmark points with a significance of $2\sigma$ or higher.

The CSI model predicts $\kappa_\lambda \simeq 1.73$, largely independent of other parameters.  
This sizable deviation can be observed at the $1\sigma$ level at both the HL-LHC and the ILC at $\sqrt{s} = 1~\mathrm{TeV}$, and at the $2\sigma$ level at the muon and $100~\mathrm{TeV}$ proton--proton colliders.

In MCHM scenarios, $\kappa_\lambda$ depends on the vacuum misalignment parameter $\xi = v^2/f^2$.  
We use $\xi = 0.1$, consistent with electroweak precision constraints~\cite{Bellazzini:2014yua}.  
In MCHM4 without higher-order corrections and in MCHM5, $\kappa_\lambda$ shows only small deviations from the SM value and remains beyond the reach of all considered colliders.  
In contrast, the parameter $C$ in MCHM4 with higher-order corrections has a significant impact on $\kappa_\lambda$. Note that the upper limits on $\xi$ have been obtained from measurements of $\kappa_{V}$ and $\kappa_{F}$ in Ref.~\cite{ATLAS:2022vkf, CMS:2022uhn} for LHC Run~2, based on Eqs.~(\ref{eq:MCHM_hVV_coupling}) and (\ref{eq:MCHM4_yukawa}). 
While it is challenging to constrain $C$ itself at the current LHC, the HL-LHC can tighten the allowed region to $-0.5 < C < 0.2$ for $\xi = 0.1$, which is improved by about an order-of-magnitude improvement from the current ones.  
For $(\xi, C) = (0.1, 0.1)$, the model becomes marginally accessible at the ILC with $\sqrt{s} = 1~\mathrm{TeV}$, the muon collider, and the $100~\mathrm{TeV}$ proton--proton collider with $1\sigma$ sensitivity.

The tadpole-induced model predicts $\kappa_\lambda = 0$ at tree level and a large negative value at the one-loop level due to top-quark contributions.  
This large deviation allows the HL-LHC to probe the model at the $1\sigma$ C.L.. 
Future colliders such as the muon collider and the $100~\mathrm{TeV}$ proton--proton collider will be able to test it at the $2\sigma$ level or higher.

To summarize the expected sensitivities at future collider experiments, Table~\ref{tab:detectability} presents the detectability of deviations in $\kappa_\lambda$ for the benchmark points considered in our analysis. 
The results are classified into three categories: ``++'' (detectable at or above the $2\sigma$ level), ``+'' (detectable at the $1\sigma$ level), and ``--'' (not detectable within the projected sensitivities). 
This table illustrates that future colliders will be necessary for probing and discriminating the extended Higgs potentials through precise measurements of the trilinear Higgs boson coupling.

\begin{table}[t]
    \centering
    \caption{\footnotesize
    Ratio of the trilinear Higgs coupling $\kappa_\lambda$ in various extended Higgs models and their detectability at future colliders.
    Detection significance is categorized as ``++'' (detectable at $\ge 2\sigma$), ``+'' (detectable at $1\sigma$), and ``--'' (not detectable).
    }
    \label{tab:detectability}
    \begin{ruledtabular}
    \begin{tabular}{lcccc}
        \textbf{Model}  & \textbf{HL-LHC} & \textbf{ILC (1~TeV)} & \textbf{Muon Col.} & \textbf{pp Col. (100~TeV)} \\
        \hline
        SMEFT ($D = 10^{-7}~\mathrm{TeV}^{-2}$) & -- & + & + & ++ \\
        CSI model                               & + & + & ++ & ++ \\
        MCHM4$_{(0)}$ ($\xi = 0.1$)            & -- & -- & -- & -- \\
        MCHM4$_{(1)}$ ($\xi = 0.1$, $C = 0.1$) & -- & + & + & + \\
        MCHM5 ($\xi = 0.1$)                    & -- & -- & -- & -- \\
        Tadpole-induced model                   & + & ++ & ++ & ++ \\
    \end{tabular}
    \end{ruledtabular}
\end{table}

%%%%%%%%%%%%%%%%%%%%%%%%%%%%%%%%%%%%%%%%%%%%%%%%%%%%%%%%%%%%%%%%%%%%
\section{Summary}
\label{sec:summary}
%%%%%%%%%%%%%%%%%%%%%%%%%%%%%%%%%%%%%%%%%%%%%%%%%%%%%%%%%%%%%%%%%%%%

In this work, we have discussed the prospects for probing various extensions of the Higgs sector of the Standard Model with future collider experiments.  
By classifying models according to the functional forms of their Higgs potentials, future collider measurements can efficiently discriminate among primary scenarios.  
We have included the one-loop corrections to the trilinear Higgs boson coupling in several frameworks: the naHEFT with dimension-six operators, the CSI model, the MCHMs, and the tadpole-induced models.  
As demonstrated in this paper, the tadpole-induced model and the CSI model can be discriminated at the 68\% C.L. by the HL-LHC and the ILC at $\sqrt{s} = 1~\mathrm{TeV}$, respectively.  
In contrast, the Higgs EFT framework with dimension-six operators and the pNGB models remain challenging to probe with the currently planned collider experiments.

For these classes of models, gravitational-wave observations offer powerful and complementary probes.
Future space-based gravitational-wave observatories such as LISA and DECIGO will be capable of detecting signatures associated with electroweak phase transitions predicted in these models, which could provide important indications of the shapes of the Higgs potentials.
Extending this framework to include the quartic Higgs boson coupling would further help clarify functional forms of various extended Higgs potentials.
However, a detailed investigation of these aspects lies beyond the scope of this paper and is left for future work.

%%%%%%%%%%%%%%%%%%%%%%%%%%%%%%%%%%%%%%%%%%%%%%%%%%%%%%%%%%%%%%%%%%%%
\section*{Acknowledgement}
%%%%%%%%%%%%%%%%%%%%%%%%%%%%%%%%%%%%%%%%%%%%%%%%%%%%%%%%%%%%%%%%%%%%

We gratefully thank Kei Yagyu and Shin Suzuki for helpful comments and discussions. 
This work was supported in part by the JSPS KAKENHI Grant Number 22K14035 [NH],  20H05852 [NH], 21K03571 [MK], and 20H00160 [MK].
The work of N.H is also supported in part by the MEXT Leading Initiative for Excellent Young Researchers Grant Number 2023L0013.  
This work was supported by JST SPRING, Grant Number JPMJSP2145 [SO].

%%%%%%%%%%%%%%%%%%%%%%%%%%%%%%%%%%
%%%%%%%%%%% References %%%%%%%%%%%
%%%%%%%%%%%%%%%%%%%%%%%%%%%%%%%%%%
\bibliographystyle{apsrev4-1}
\bibliography{references}

\end{document}